\documentclass[fleqn,usenatbib]{mnras}

\usepackage{newtxtext,newtxmath}

\usepackage[T1]{fontenc}

\DeclareRobustCommand{\VAN}[3]{#2}
\let\VANthebibliography\thebibliography
\def\thebibliography{\DeclareRobustCommand{\VAN}[3]{##3}\VANthebibliography}

\usepackage{graphicx}	
\usepackage{amsmath}	
\usepackage{color}
\usepackage{multirow}
\usepackage{float}
\usepackage{subfigure}
\usepackage{booktabs}
\usepackage[utf8]{inputenc}
\usepackage[acronym]{glossaries}
\usepackage{ulem}

\newcommand{\msun}{\ensuremath{\mathrm{M}_{\rm \odot}}}

\newcommand{\eorb}{E_{\rm orbital}}
\newcommand{\jsref}{\ensuremath{D_{{\rm JS, ref}}}}
\newcommand{\jsnoise}{\ensuremath{D_{{\rm JS, noise}}}}
\newcommand{\jsrec}{\ensuremath{D_{{\rm JS, recover}}}}

\newacronym{bbh}{BBH}{binary black hole}
\newacronym{bh}{BH}{black hole}
\newacronym{gw}{GW}{gravitational wave}

\title[Binary black hole mimics]{Mimicking Mergers: Mistaking Black Hole Captures as Mergers}

\author[Weichangfeng Guo et al.]{%
Weichangfeng Guo,$^{1}$ \thanks{E-mail: gwcf@mail.bnu.edu.cn}%
Daniel Williams,$^{2}$ 
Ik Siong Heng,$^{2}$ 
Hunter Gabbard,$^{2}$ 
\newauthor
Yeong-Bok Bae,$^{3}$
Gungwon Kang,$^{4}$
and Zong-Hong Zhu,$^{1,5}$ 
\\
$^{1}$Department of Astronomy, Beijing Normal University, Beijing 100875, China\\
$^{2}$SUPA, School of Physics and Astronomy, University of Glasgow, Glasgow G12 8QQ, United Kingdom\\
$^{3}$Center for Theoretical Physics of the Universe, Institute for Basic Science (IBS), Daejeon 34126, Korea\\
$^{4}$Department of Physics, Chung-Ang University, Seoul 06974, Korea\\
$^{5}$School of Physics and Technology, Wuhan University, Wuhan 430072, China\\
}

\date{Accepted 2022 August 15. Received 2022 July 30; in original form 2022 March 30}

\pubyear{2022}
\begin{document}
\label{firstpage}
\pagerange{\pageref{firstpage}--\pageref{lastpage}}
\maketitle

\begin{abstract}
As the number of gravitational wave observations has increased in recent years, the variety of sources has broadened.
Here we investigate whether it is possible for the current generation of detectors to distinguish between very short-lived gravitational wave signals from mergers between high-mass black holes, and the signal produced by a close encounter between two black holes which results in gravitational capture, and ultimately a merger.
We compare the posterior probability distributions produced by analysing simulated signals from both types of progenitor events, both under ideal and realistic scenarios. 
We show that while, under ideal conditions it is possible to distinguish both progenitors, under more realistic conditions they are indistinguishable.
This has important implications for the interpretation of such short signals, and we therefore advocate that these signals be the focus of additional investigation even when satisfactory results have been achieved from standard analyses.

\end{abstract}

\section{Introduction}
\label{sec:intro} 
In recent years Binary black hole (BBH) observations have become a mainstay of gravitational wave (GW)~\citep{gw} detection~\citep{gwtc-1,gwtc-2,gwtc-3}.
Observable BBH signals are produced during the late stages of the decay of a bound orbit of two black holes (BHs): these observations are typically short-lived, and have a distinctive morphology, produced by the final orbits (the ``inspiral'' phase), the merger of the two BHs, and finally the ``ringdown'' of the final, merged BH.
Most BBHs are expected to be circularized before the merging due to the loss of the orbital energies during the inspiral phase~\citep{circular_1,circular_2,Abadie_2010}, therefore the current analyses of GW data and the parameter estimations of GW sources have focused on binaries with circular orbits. 
Eccentricity has not been detected~\citep{burst_search_2,eccentricity_search_2,eccentricity_search_3} in the O1 and O2 observing runs of LIGO/Virgo~\citep{ligo,virgo}, while a high-mass BBH event in O3a, GW190521~\citep{gw190521,gw190521_2}, shows evidence it may have been both highly eccentric and dynamically formed, as a result of parameter estimation analysis~\citep{gw190521_dynamical,gw190521_dynamical_2} using both spin-aligned eccentric waveform approximants, \texttt{SEOBNRE}~\citep{gw190521_dynamical,SEOBNRE_1,SEOBNRE_2} and \texttt{TEOBResumS}~\citep{gw190521_dynamical_2,TEOBResumS_2}; and numerical numerical relativity simulations~\citep{gw190521_dynamical_3}. 

There are several dynamical-formation scenarios~\citep{dynamical_1,iso_spins_1,dynamical_3,dynamical_4,dynamical_5,dynamical_6,dynamical_7} which support the formation and merger of binary systems while retaining eccentricity throughout their lifetime. 
One of the formation processes of these eccentric BH binaries is a gravitational radiation driven capture. 
Captures are expected to take place in galactic nuclei, where the central super massive BH creates a steep density cusp of stellar-mass BHs.
This can provide a suitable environment for eccentric BBH formation~\citep{bhencounter}.
In accordance with the conditions of the initial orbital energy $E_{\rm orbital}$ and the initial angular momentum $L_{\rm initial}$, the motion of the binary systems can be classified as bound (circular or elliptical) and unbound (parabolic or hyperbolic) orbits~\citep{hyperbolic,encounter_gw_production_1,encounter_gw_production_2,parabolic}. 

Gravitationally unbound interactions, or encounters, occur when $\eorb \geq 0$ without direct capture, and the trajectory of one component of the system will be either parabolic or hyperbolic relative to the other.
The GW signals produced from such encounters do not resemble those from bound systems, but will instead produce a strong burst of radiation as the objects have their closest encounter.
The process may be considered a gravitational analogy to Bremsstrahlung~\citep{gw_brem_1}.

In this work we will focus on parabolic encounters~\citep{parabolic_waveform} with $E_{\rm orbital}=0$. 
BH encounters that correspond to parabolic orbits at infinity merge quickly if their initial angular momentum is not suitably large enough~\citep{bhencounter}. We use the term ``parabolic black hole capture'' to refer to the systems which form and merge in this way.
In other cases, the binaries tentatively pass by, and finally merge in the very distant future.

We expect the GW signal from BH capture to be detectable both in the current generation of detectors, and planned future detectors~\citep{bh_encounter_detection_1,bh_encounter_detection_2}. 
The expected event rate for hyperbolic BH encounters in galactic nuclei is comparable to estimates for other sources of GW signals, which is $\sim 0.9\ {\rm Gpc}^{-3}\ {\rm yr}^{-1} $~\citep{bh_encounter_detection_3}, independent of any detector. 

GW signals from BBH events have a characteristic morphology: during the inspiral, before the two BHs merge the signal is sinusoidal, with a growing frequency as the radius of the binary decreases. 
The remainder of the signal is produced from the merger, and the ringdown. 
As the total mass of the binary increases the frequency at which the merger occurs decreases, and consequently, for high mass systems the merger may occur close to the lowest frequency the detector is capable of successfully observing. 
In such a scenario the signal might appear to have little or no inspiral due to the lower frequency limit of ground-based detectors.
The waveforms of BH encounters, or parabolic BH captures will often have no inspiral, or only a small number of cycles of inspiral. 
As a result it is not implausible that an event may be misinterpretted as a high-mass BH coalescence when in fact it had been a BH capture.
An example of an observation which fits these conditions is GW190521, which has been interpreted as a BBH coalescence with total mass around $142\ \msun$~\citep{gw190521,gw190521_2,gw190521_4,gw190521_5,gw190521_3}.

The use of unmodelled ``burst'' searches has been proposed for these highly eccentric mergers~\citep{burst_search,burst_search_2,burst_search_3}, though they are less capable of digging deep into the noise for signals and their sensitivity is hard to quantify~\citep{search_comment}. 
Therefore it is probable that GW190521-like signals will be detected using analyses designed to identify and analyse BBH signals in the future, rather than those designed for more exotic waveform morphologies. 
At present parameter estimation analyses using an approximant for parabolic capture signals is not viable, due to a lack of sufficiently flexible waveform models.
We therefore ask if analyses using BBH waveform models will produce different, and distinguishable, results when used to analyse both BBH and parabolic capture signals.
In order to do this we conducted two sets of analyses: one analysing simulated BBH signals injected into simulated noise, and one analysing simulated parabolic capture signals in simulated noise.
We then compare the posterior probability distributions of the BBH and parabolic capture injections.

The conventional approach for parameter estimation used in GW analysis uses Bayesian inference and stochastic sampling, which usually requires a large amount of computation.
Therefore, a neural network that can quickly give a posterior under BBH model while maintaining a fairly high accuracy was employed. Comparative studies of such a neural network approach and conventional Bayesian parameter estimation techniques have shown that a properly trained neural network is capable of emulating a posterior distribution to an acceptable level of accuracy~\citep{vitamin}.
Given the large amount of simulated data required for our study, we used the network to obtain the posteriors for each simulation. 
We compare the mass, distance and merger time posterior distributions using the Jensen-Shannon (JS) divergence~\citep{jsd_1} to quantify any differences between these posteriors for the simulated population of black hole captures and BBH signals.

The outline of this paper is as follows. 
In section \ref{sec:data}, we describe the production of the simulated parabolic BH capture and BBH signals which are used for the analysis. 
In section \ref{sec:methods}, describe the parameter estimation method, and the use of a deep learning approach to improve computing speed. 
The results of this analysis are presented in section \ref{sec:result}. 
Finally, the main outcomes of the work and possible future directions are summarised in section \ref{sec:conclusion}.

\section{Mock data creation}
\label{sec:data}

Recent advances in numerical relativity~\citep{parabolic_waveform} have allowed the production of gravitational waveforms for unequal mass BH encounters under the parabolic approximation, which were used in this work.
The generated waveforms are applicable for non-spinning pairs of BHs with relative velocity up to $10\sim20\ \%$ of the speed of light. 
The four waveforms we used to produce our mock data are from parabolic BH captures with mass ratios $q \in \{ 1, 4, 8, 16\}$. 
While these waveforms contain a merger and a ringdown, they lack the characteristic sinusoidal morphology of the inspiral.

\begin{figure*} 
	\centering  
	\vspace{0.35cm} 
	\subfigtopskip=2pt 
	\subfigbottomskip=0.5pt 
	\subfigcapskip=-5pt 
	\subfigure[]{
		\label{plt:noisefree_bbh}
		\includegraphics[width=\columnwidth]{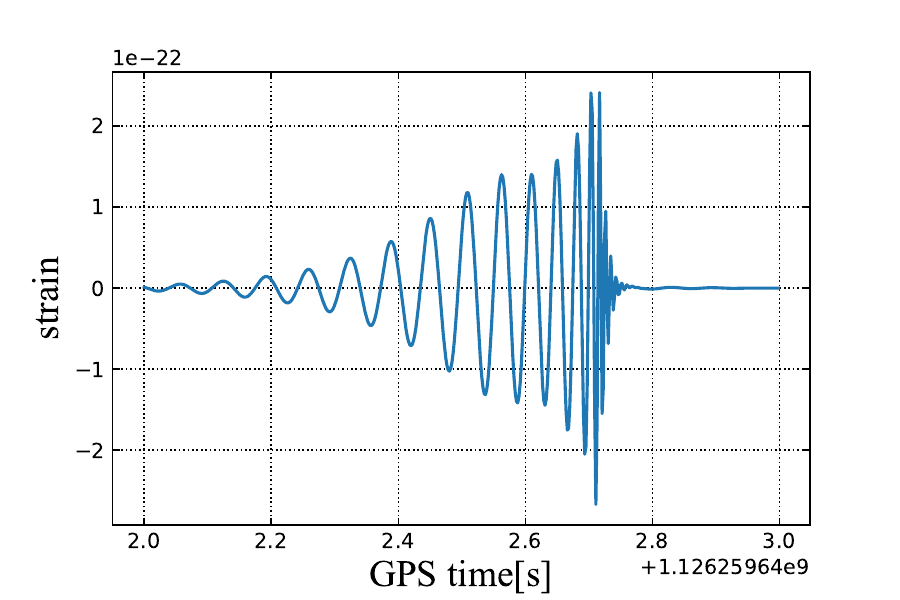}}
	\subfigure[]{
		\label{plt:noisy_bbh}
		\includegraphics[width=\columnwidth]{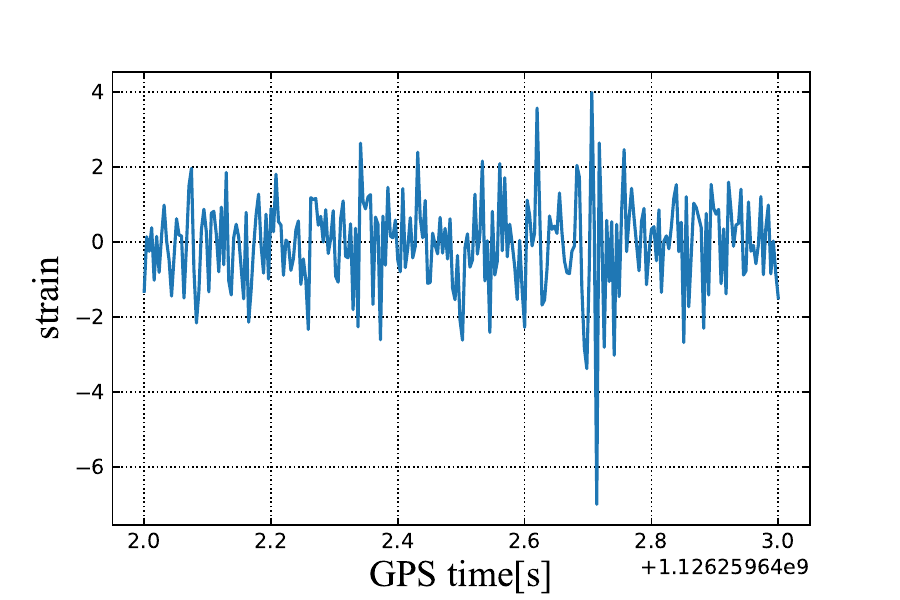}}
	\subfigure[]{
		\label{plt:noisefree_be}
		\includegraphics[width=\columnwidth]{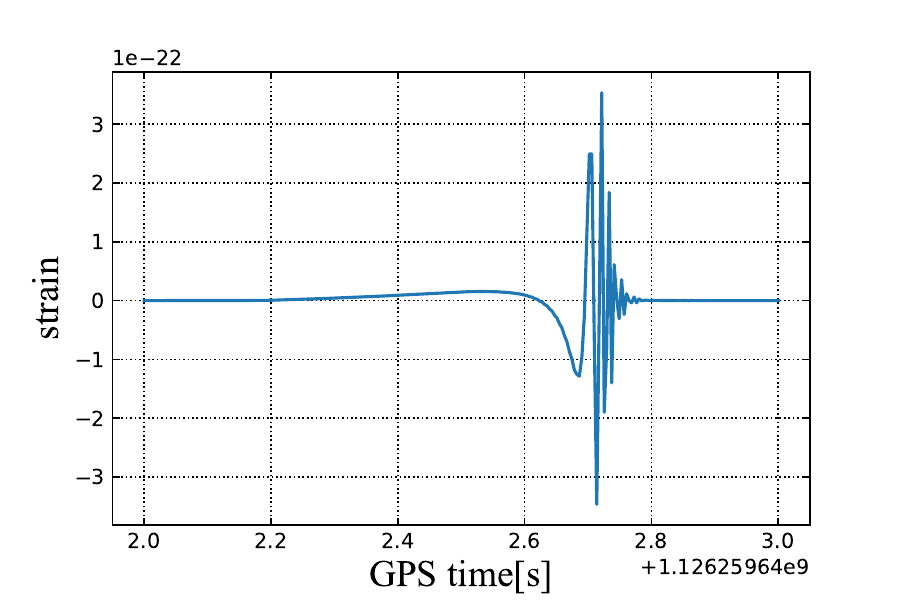}}
	\subfigure[]{
		\label{plt:noisy_be}
		\includegraphics[width=\columnwidth]{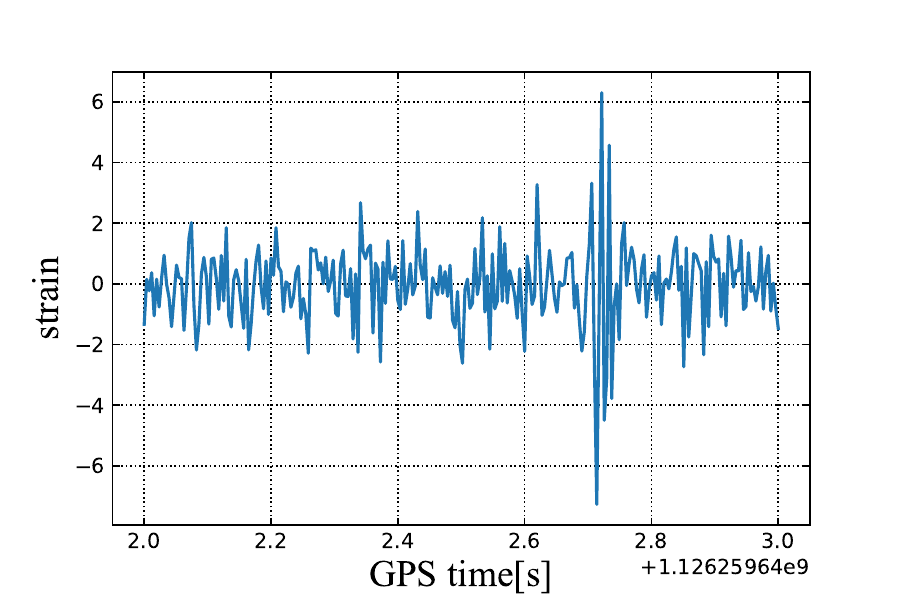}}
	\caption{Examples of BBH and parabolic BH capture injections which used in this work. The panels on the left show the signal before injection, and those on the right the signal whitened against the power spectrum of the simulated detector noise. 
	In the top row (a) represents a typical high-mass BBH signal for a system with $m_{1}=78\ \msun$, $m_{2}=72\ \msun$ at a distance of $1400\ {\rm Mpc}$; (b) depicts the signal from (a) whitened.
	In the bottom row (c) represents a parabolic BH capture signal for a system with mass ratio $q = 1$, and a total mass of $150\ \msun$ at a distance of $5000\ {\rm Mpc}$, with (d) depicting it whitened. All waveforms in this paper correspond to a three-detector configuration, while here we only show the signal in H1 detector.}
	\label{plt:wfm}
\end{figure*}

We used the \texttt{Minke}~\citep{minke} python package to generate the parabolic capture injections.
\texttt{Minke} is a toolkit designed to produce injection sets for signals derived from numerical relativity simulations, and performs the appropriate rescaling required to produce waveforms for systems with any total mass, and at any luminosity distance.
The signal is then convolved with the detector's antenna pattern, and time shifted for the corresponding detector.
The process for producing the mock data set using \texttt{Minke} is as follows.

For each of the four mass ratios we considered we chose a luminosity distance and a sky location for the parabolic capture waveforms which produced a posterior probability distribution when analysed by \texttt{VItamin} which was visually similar to the posterior from analysing a high-mass BBH signal. The posterior of a typical BBH has a shape with the following features: it should have obvious peaks and narrow width of the marginal distributions, which indicates that the corresponding parameters have been inferred well under the BBH model.
The maxima of the marginal distributions from BH captures are not required to be in the same locations as a typical BBH.
The visual method essentially checks that the posterior distributions neither rail against, nor are compressed towards one of the edges of the prior for a particular parameter.
This led us choosing the waveform parameters shown in Table~\ref{tab:injection}. 
The simulated data are created with a fixed total mass of $150\ \msun$, and a luminosity distance $d_0$ in $[100,8000]\ {\rm Mpc}$. 
Here, and elsewhere in this work, masses are quoted in the detector frame.
The right ascension and declination, $\alpha$, $\delta$, and the waveform polarisation, $\psi$ were distributed uniformly for all waveforms. 
The detectors we used are LIGO Handford (H1), LIGO Livingston (L1)~\citep{ligo}, and Virgo (V1)~\citep{virgo}.

\begin{table}
\centering
\begin{tabular}[t]{lcc}
\toprule
Parameter & Injection\\
\hline
$m_{\rm total}$ (M$_\odot$) & \multicolumn{1}{c}{150} \\
$d_{\rm L}$ (Mpc) & \multicolumn{1}{c}{$d_{0}$} \\
$t_{0}$ (s) & \multicolumn{1}{c}{0.22}\\
$\alpha$ & [0, $2\pi$]\\
$\delta$ & [$-\pi/2$, $\pi/2$]\\
$\psi$ & [0, $2\pi$]\\
\hline
duration (s)& \multicolumn{1}{c}{1} \\
$\rm t_{start}$ (GPS time)& \multicolumn{1}{c}{1126259642} \\
$\rm t_{ref}$ (GPS time)& \multicolumn{1}{c}{1126259642.5} \\
detector network  & \multicolumn{1}{c}{H1, L1, V1} \\
\bottomrule
\end{tabular}
\caption{
The injections of parabolic BH capture mock data used for \texttt{VItamin} analysis. 
We list the start time $t_{\rm start}$, the reference time $t_{\rm ref}$ in GPS time and the fixed merger time $t_0$ = $0.22\ s$, where the merger time in GPS time $t_{\rm merger}$ = $t_{\rm ref}$ + $t_0$. 
Here, and elsewhere in this work, masses are quoted in the detector frame.
}
\label{tab:injection}
\end{table}

The start time $t_{\rm start}$ was specified when generating mock signal, while the merger time $t_0$ and signal length could not be. 
For the parabolic BH capture waveforms with a mass ratio of $1$, $4$, $8$, $16$, the length of the raw data produced by \texttt{Minke} ranges from $0.88\ {\rm s}$ to $1.34\ {\rm s}$.
We manually truncated the timeseries, or padded it with zeros, to fit the one-second analysis constraint of \texttt{VItamin}. 
At the same time, we put the signal's peak at $t_0 = 0.22\ {\rm s}$, which lies within the pre-trained prior range $[0.15, 0.35]\ {\rm s}$ expected by \texttt{VItamin}.  
To make the network function properly, signals were processed in the same way as the training data. 
\texttt{VItamin} requires that the input data is whitened using detector amplitude spectral density (ASD), and given a zero mean, unit variance Gaussian noise. This whitening process was adopted mainly to scale the data more properly for input to the neural network, and we employed aLIGO zero detuning\footnote{See \url{https://dcc.ligo.org/T1800044-v5/public}}, high power design sensitivity ASD for H1 and L1, and the advanced Virgo\footnote{See \url{https://dcc.ligo.org/LIGO-P1200087-v42/public}} ASD for V1. Examples of BBH and parabolic BH capture injections are showed in Figure~\ref{plt:wfm}.

\section{Analysis}
\label{sec:methods} 

The most widely used approach to analysing BBH signals uses Bayesian inference~\citep{bayes_1,bayes_2}. 
One analysis pipeline which is widely used is \texttt{Bilby}~\citep{bilby,bilby_2}, a modular python package. However, it is computationally intensive because it uses stochastic sampling techniques to estimate the posterior.
Instead, we used \texttt{VItamin}~\citep{vitamin} to produce rapid posterior estimates for each injection.
Due to the high-mass systems we focus on, we first used the pre-training function supported by \texttt{VItamin} to expand its prior parameter space to Table~\ref{tab:prior_vtm}.
Once we obtained a posterior that is visually similar to that for a BBH, we applied Bayesian inference to the corresponding signal using non-spinning and spinning BBH templates. For the two inferences, we employed the \texttt{IMRPhenomPv2}~\citep{imr_template,imr_template_present} approximant respectively\footnote{ \texttt{IMRPhenomPv2} has 6 parameters to model the spins of BBH system. In order to produce non-spinning waveforms from it we set the 6 spins as zero.} 
because it was used to train \texttt{VItamin}.
We carried-out these analyses using \texttt{Bilby} as corroboration for \texttt{VItamin} analysis.

\subsection{Bayesian Inference}
\label{sec:bayes} 

The probability distribution on a set of parameters, conditional on the measured data, can be determined using Bayes Theorem, which can be represented as
\begin{equation}\label{eq:bayes_theorem} 
p(x|y) = \frac{p(y|x) p(x)}{p(y)}, 
\end{equation}
where $x$ are the parameters, $y$ are the observed data, $p(x)$ is the prior on the parameters, $p(y)$ is the probability of the data, $p(x|y)$ is the posterior, and $p(y|x)$ is the likelihood. 

GW parameter estimation analyses typically require the exploration of a very large parameter space, while analysing a large volume of data.
To address this it is typical to use a stochastic sampler to reconstruct the posterior. 
This sampling can be done with a variety of techniques, including Nested Sampling and Markov chain Monte Carlo~\citep{montecarlo_1,mcmc_application_1,mcmc_application_2} methods. 
The popular software packages used by LIGO parameter estimation analyses are \texttt{LALInference} and \texttt{Bilby}, which offer multiple sampling methods.
We used \texttt{Bilby}, a Bayesian inference library for GW astronomy, as an interface for the \texttt{dynesty}~\citep{dynesty} sampler. 

Once the appropriate posteriors had been obtained in section~\ref{sec:vitamin}, one example of the corresponding signals was expanded with a data segment of 4 seconds and a sampling rate of $1024\ {\rm Hz}$ for a precise analysis. We used the \texttt{dynesty} sampler, and both spinning and non-spinning BBH waveforms drawn from the \texttt{IMRPhenomPv2} model to perform parameter estimation on the data. The priors we used are shown in Table~\ref{tab:prior_non-spinnning} and Table~\ref{tab:prior_spinning}.

We then calculated the Bayes factors $K$ for non-spinning BBH-template and spinning BBH-template against the noise. 
The Bayes factor is defined as 
\begin{equation}\label{eq:bayes_factor} 
K = \frac{p(y|x,H_1)}{p(y|x,H_2)} = \frac{\int p(x_1|H_1)p(y|x_1,H_1)dx_1}{\int p(x_2|H_2)p(y|x_2,H_2)dx_2},
\end{equation}
where $H_1$, $H_2$ are two different hypotheses.
With a $K$ \textgreater 1 indicating greater support for $H_1$ hypothesis.

Finally, we used the median recovered values from the parabolic capture signal posterior to create injections using two BBH models (one spinning, and one non-spinning).
These two recovered signals were compared with the corresponding parabolic capture, demonstrating the ability of a parabolic capture to mimic a high-mass BH when analysed with the \texttt{IMRPhenomPv2} waveform approximant.

\subsection{VItamin analysis}
\label{sec:vitamin}

\texttt{VItamin} is a recently proposed network for BBH signals based on a conditional variational autoencoder~\citep{CAVE_1,CAVE_2}, which has been shown to produce samples describing the posterior distribution six orders of magnitude faster than the traditional Bayesian approach~\citep{vitamin}. 
The network used non-spinning BBH approximant \texttt{IMRPhenomPv2}. 
It omits the six additional parameters required to model the spins of the BBH system and produces posteriors on eight parameters: the component masses $m_1$, $m_2$, the luminosity distance $d_{\rm L}$, the time of coalescence $t_0$, the binary inclination $\theta_{jn}$, right ascension $\alpha$, and declination $\delta$. 
The phase at coalescence $\phi_0$ and the GW polarisation angle $\psi$ are internally marginalized out.
For each parameter we used a uniform prior, with the exception of the declination and inclination parameters for which we used priors which were uniform in $\cos (\delta)$ and $\sin\theta_{jn}$.
The corresponding prior ranges are defined in Table~\ref{tab:prior_vtm}.
The initial prior range of \texttt{VItamin} focuses on low-mass BH binaries, and the upper limit of the component mass is $80\ \msun$. 
However, we trained the network with a customized prior, increasing the maximum component mass to $160\ \msun$ to deal with the high-mass BBH system in this study.
The BBH signals used as training and test data were produced using \texttt{IMRPhenomPv2}, with a minimum cutoff frequency of $20\ {\rm Hz}$. 
The training procedure only needed to be performed once, and took $\mathcal{O}(1)$ day to complete.
The resulting trained network could then quickly generate samples describing the posterior distribution, which was proved to achieve the same accuracy of results as trusted benchmark analyses used within the LIGO-Virgo Collaboration~\citep{vitamin}. 
For BBH signals, GW data is usually sampled to frequencies between $1$ and $16\,\mathrm{kHz}$, depending upon the mass of binary. 
We have chosen a low sampling rate of $256\ {\rm Hz}$ for the \texttt{VItamin} network, in order to decrease the computational time required to train it. 
We observed that, the main frequency component of BH capture signals with total mass of $150\ \msun$ is around 80 Hz, which is below Nyquist sampling rate $f_{\rm Nyquist}=128\ {\rm Hz}$, thus the signal is well covered by the sampling range.

\begin{table} 
\centering

\begin{tabular}[t]{lccc}
\toprule
Parameter & min & max & prior\\
\hline
$m_{1, 2}$ (M$_\odot$) & 30 & 160 & uniform\\
$d_{\rm L}$ (Mpc) & 1000 & 3000 & uniform\\
$t_{0}$ (s) & 0.15 & 0.35 & uniform\\
$\alpha$ & 0 & $2\pi$ & uniform\\
$\delta$ & $-\pi/2$ & $\pi/2$ & cosine\\
$\theta_{jn}$ & 0 & $\pi$ & sine\\

\hline
spins & \multicolumn{2}{c}{0} & - \\
duration (s)& \multicolumn{2}{c}{1} & - \\
$\rm t_{start}$ (GPS time)& \multicolumn{2}{c}{1126259642.0} & - \\
$\rm t_{ref}$ (GPS time)& \multicolumn{2}{c}{1126259642.5} & - \\
detector network & \multicolumn{3}{c}{H1, L1, V1} \\
\bottomrule
\end{tabular}
\caption{The priors and fixed parameter values used on non-spinning BBH model parameters for \texttt{VItamin} analysis.}
\label{tab:prior_vtm}
\end{table}

For all parabolic BH capture waveforms, we used one set of sky location injections ($\alpha$, $\delta$, and $\psi$) that contains 100 samples. 
We only adjusted $d_{\rm L}$ injection, for the appropriate posteriors, which were given by \texttt{VItamin} network rapidly once the data had been input.
$d_{\rm L}$ injection is a critical factor due to the effect of waveform scaling.
It must be chosen such that the injected waveform has an SNR which would be detectable, and produces a plausible posterior distribution which might be mistaken as that of a high-mass BBH.

\begin{table}\label{ns_prior} 
\centering

\begin{tabular}[t]{lccc}
\toprule
Parameter & min & max & prior\\
\hline
$m_{1, 2}$ (M$_\odot$)  & 30 & 160 & uniform\\
$d_{\rm L}$ (Mpc)  & 1000 & 3000 & uniform\\
$t_{0}$ (s) & 0.15 & 0.35 & uniform\\
$\alpha$ & 0 & $2\pi$ & uniform\\
$\delta$ & $-\pi/2$ & $\pi/2$ & cosine\\
$\theta_{jn}$ & 0 & $\pi$ & sine\\
$\psi$ &  0  & $\pi$ & uniform\\
$\phi$ &  0  & $2\pi$ & uniform\\
\hline
spins & \multicolumn{2}{c}{0} & - \\
duration (s)& \multicolumn{2}{c}{4} & - \\
$\rm t_{start}$ (GPS time)& \multicolumn{2}{c}{1126259642.0} & - \\
$\rm t_{ref}$ (GPS time)& \multicolumn{2}{c}{1126259644.5}  & - \\
detector network & \multicolumn{3}{c}{H1, L1, V1} \\
\bottomrule
\end{tabular}
\caption{The priors and fixed parameter values used on non-spinning BBH model parameters for \texttt{Bilby} analysis. In this analysis we use a 4-second duration timeseries.}
\label{tab:prior_non-spinnning}
\end{table}

\begin{table}\label{s_prior} 
\centering

\begin{tabular}[t]{lccc}
\toprule
Parameter & min & max & prior\\
\hline
$m_{1, 2}$ (M$_\odot$) & 30 & 160 & uniform\\
$d_{\rm L}$ (Mpc) & 1000 & 3000 & uniform\\
$t_{0}$ (s) & 0.15 & 0.35 &  uniform\\
$\alpha$ & 0 & $2\pi$ &  uniform\\
$\delta$ & $-\pi/2$ & $\pi/2$ &  cosine\\
$\theta_{jn}$ & 0 & $\pi$ &  sine\\
$\psi$ &  0  & $\pi$ &  uniform\\
$\phi$ &  0  & $2\pi$ &  uniform\\
$a_{1,2}$ & 0 & 0.99 &  uniform\\
$\theta_{1,2}$ & 0 & $\pi$ &  sine\\
$\Delta\phi$ &  0  & $2\pi$ &  uniform\\
$\phi_{JL}$ &  0  & $2\pi$ &  uniform\\
\hline
duration (s)& \multicolumn{2}{c}{4} & - \\
$\rm t_{start}$ (GPS time)& \multicolumn{2}{c}{1126259642.0} & - \\
$\rm t_{ref}$ (GPS time)& \multicolumn{2}{c}{1126259644.5}  & - \\
detector network & \multicolumn{3}{c}{H1, L1, V1} \\
\bottomrule
\end{tabular}
\caption{The priors and fixed parameter values used on spinning BBH model parameters for \texttt{Bilby} analysis. In this analysis we use a 4-second duration timeseries.}
\label{tab:prior_spinning}
\end{table}

\section{Results}
\label{sec:result}

In the \texttt{VItamin} recovery study on parabolic BH capture using non-spinning BBH approximate \texttt{IMRPhenomPv2}, we did obtain the posteriors look superficially like a BBH posterior.
\footnote{
The posterior probability distribution of a parabolic BH capture which mimics a BBH determined by \texttt{VItamin} could be seen in Figure~\ref{vtm_be}. 
As a comparison, we also display the \texttt{VItamin} posterior of a typical high-mass BBH in Figure~\ref{vtm_bbh}.} 
The $d_{\rm L}$ injections to produce such posteriors are recorded in Table~\ref{tab:search_result}. 
We calculated the optimal SNR in each detector, which is defined as
\begin{equation}
    \label{eq:snr}
    \rho_{\rm opt} = 2{\left [\int_{f_{\min}}^{f_{\max}}\frac{|h^{2}(f)|}{S_{h}(f)}\ {\rm d}f \right ]^{\frac{1}{2}}},
\end{equation}

where $h(f)$ is the Fourier transform of the (time-domain) GW signal, $S_{h}(f)$ is the one-sided noise spectral density in units of $\rm Hz^{-1}$,
and $f_{\min}\leq f \leq f_{\max}$ correspond to the frequency band of the instrument. 
We found that, in the step of adjusting $d_{\rm L}$ injection, waveforms with a higher mass ratio produced a detectable SNR to smaller luminosity distances. 

\begin{figure*} 
	\centering  
    \includegraphics[width=\textwidth]{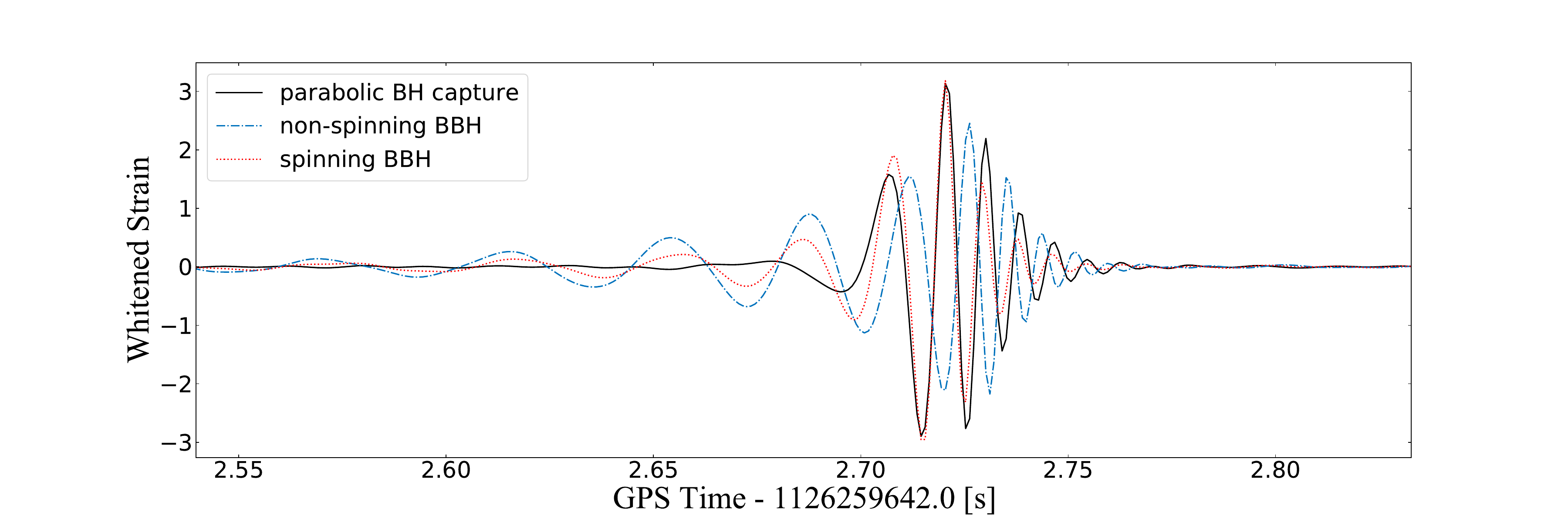}
	\caption{The whitened parabolic BH capture and the whitened recovered BBH signals at detector H1. The waveform with mass ratio $q$ = 1 was injected with a total mass of 150 \msun at a distance of 5000 Mpc. The parameter estimation was then performed on the signal using non-spinning and spinning BBH model \texttt{IMRPhenomPv2} by \texttt{Bilby}. 
	Here we present the waveform corresponding to the median values of each parameter's posterior distribution. }
	\label{signal_comparison}
\end{figure*}

We also conducted parameter estimation using \texttt{Bilby} on one parabolic capture signal, with the posterior distribution sampled by \texttt{dynesty}.
\footnote{The posterior probability distributions for a parabolic capture signal analysed by \texttt{Bilby} is shown in Figure~\ref{blb_be} and \ref{blb_be_spin} using non-spinning and spinning models respectively.}
Both spinning and non-spinning BBH templates \texttt{IMRPhenomPv2} have a high log Bayes factor $\ln{K} = 134.620^{\pm 0.168}, 117.134^{\pm 0.155}$ against noise, strongly supporting the hypothesis that the signal is a BBH merger. 
As a comparison, the reference BBH data gives a log Bayes factor $\ln{K} =157.363^{\pm0.209}$ compared to the noise hypothesis in the non-spinning BBH analysis.
Then we have the log Bayes factor between spinning and non-spinning BBH templates $\ln{K} = 17.486^{\pm 0.229}$, which is quite small. 
This illustrates that it is very difficult for the Bayes factor to distinguish between the spinning and non-spinning models in this case. %
We then generated the signal corresponding to the median values of each waveform parameter's posterior, and compared it with the original signal from Figure~\ref{signal_comparison}, and we show the whitened waveforms of both the parabolic BH capture signal and the recovered BBH signal overlayed.
Here we can see the strong similarty in the merger-ringdown phase.
\begin{table}
\centering

\begin{tabular}{lll}
\toprule
mass ratio $q$ & $d_{\rm L}$ injection (Mpc)\\ \hline
1     & 5000  \\
4     & 2000 \\
8     & 1500 \\
16    & 500 \\
\bottomrule
\end{tabular}
\caption{The $d_{\rm L}$ injections to produce $\texttt{VItamin}$ posteriors that are visually similar to BBH.
It decreases with the increase in the mass ratio $q$. }
\label{tab:search_result}
\end{table}

Furthermore, we might be able to make a case that the resemblance is more than superficial by reference to a statistic. 
To do this we calculate the JS divergence~\citep{jsd_1} between the posterior distributions calculated by analysing both injected BBH signals and injected parabolic capture waveforms.
If the posterior distributions from an injected BBH and an parabolic capture signal are not statistically distinctive they will have a small JS divergence, and we can infer that the use of the incorrect waveform model in the analysis would not be detected.
The JS divergence is a symmetrised and smoothed measure of the distance between two probability distributions $p(x)$ and $q(x)$ defined as

\begin{equation}
D_{\rm JS}(p\mid q)=\frac{1}{2}\left [ D_{\rm KL}(p\mid s) +D_{\rm KL}(q\mid s)\right ],
\end{equation}

where $s=1/2(p+q)$ and $D_{KL}$ is the Kullback-Leibler divergence between the distributions $p(x)$ and $q(x)$ expressed as

\begin{equation}
D_{\rm KL}(p\mid q)=\int p(x) \log_{2}\left (\frac{p(x)}{q(x)}  \right )dx.  
\end{equation}

JS divergence ranges between [0, 1], a greater value of which indicates that the posteriors from two signals have a greater difference therefore they could be well distinguished.
The two JS divergences we considered are:   
\begin{enumerate}
\item $\jsnoise$: the divergence between posteriors of reference BBH signal with different white noise realisations.
\item $\jsref$: the divergence between posteriors of parabolic capture and reference BBH signal, with the same noise realisation.
\end{enumerate}
$\jsnoise$ reflects the volatility of the \texttt{VItamin} results when dealing with different white noise, whereas $\jsref$ represents the bias of the \texttt{IMRPhenomPv2} template when modelling parabolic BH capture signals. 
We expect that $\jsref$ should be obviously greater than the $\jsnoise$, in this case calculating $\jsref$ could be considered as a fast approach to distinguish BBH and parabolic BH capture events.

\begin{table*}
\centering
\begin{tabular}{lccccccccc}
\hline
mock signal &  $m_1$ (\msun) & $m_2$ (\msun) & $d_{\rm L}$ (Mpc) & $t_0$ (s) & $\alpha$ & $\delta$ &  $\psi$ & $\theta_{jn}$ & network SNR \\ \hline
parabolic BH capture m1 & 75 & 75 & 5000 & 0.22 & 0.89 & -0.94 & 1.54 & - & 11.13\\
recovered BBH & 76 & 68 & 1624 & 0.25 & 1.69 & 1.20 & - & 1.33 & 10.76\\
\hline
parabolic BH capture m4 & 120 & 30 & 2000 & 0.22 & 0.89 & -0.94 & 1.54 & - & 7.63\\
recovered BBH & 88 & 75 & 2278 & 0.26 & 4.69 & 1.21 & - & 1.76 & 4.38\\
\hline
parabolic BH capture m8 & 133.3 & 16.7 & 1500 & 0.22 & 0.89 & -0.94 & 1.54 & - & 9.93\\
recovered BBH & 98 & 83 & 1804 & 0.26 & 4.66 & 1.23 & - & 1.78 & 6.77\\
\hline
parabolic BH capture m16 & 141.2 & 8.8 & 500 & 0.22 & 0.89 & -0.94 & 1.54 & - & 13.90\\ 
recovered BBH & 104 & 90 & 1647 & 0.26 & 1.94 & 1.24 & - & 1.30 & 11.86\\ 
\hline
reference BBH & 78 & 72 & 1400 & 0.22 & 0.89 & -0.94 & 1.54 & 1.51 & 11.27\\ 
\hline
\end{tabular}
\caption{The injections of mock signals used for JS divergence analysis, including parabolic BH capture, its recovered BBH and reference BBH. For parabolic BH capture, we took the average peak value of \texttt{VItamin} posterior as the recovered injection.
The inefficiency and bias introduced by analysing the parabolic BH capture signal with a non-spinning BBH model \texttt{IMRPhenomPv2} can be seen clearly, as waveforms with a higher mass ratio were recovered to a higher total mass and lower luminosity distance with a detectable SNR. NB: $\psi$ is marginalized in \texttt{VItamin} inference, so we used $\psi=0$ for the injection. $\theta_{jn}$ is not an effective parameter for the parabolic BH capture waveform. We also note that the start time $t_{\rm start} = 1126259642.0$, the reference time $t_{\rm ref}=1126259642.5$ in GPS time, and the merger time $t_{\rm merger}$ = $t_{\rm ref}$ + $t_0$. 
}\label{tab:bias}
\end{table*}

We then created mock data for the JS divergence analysis. 
For each parabolic BH capture waveform, we reproduced one signal which can generate a posterior which is visually similar to one from a BBH.
We analysed 100 noise realisations with the same signal injected, and produced injections at the same sky location.
As a result, by considering the same injection time $t_0$, the antenna pattern is the same for each waveform. The injections and corresponding posterior peaks from the recovery are presented in Table~\ref{tab:bias}. 
For reference, we also analysed 100 noise realisations with BBH signals, where the signal parameters: total mass, right ascension $\alpha$, declination $\delta$, and merger time $t_0$, are the same as the parabolic BH capture.
$d_{\rm L}$ was changed in order to scale the BBH signal's amplitude to be similar to the parabolic BH capture by visual comparison. 
Its injections and corresponding posterior peaks from the recovery are also shown in Table~\ref{tab:bias}.

In this way, we created a situation where two similar-looking GW events, one BBH and one parabolic capture were observed.
We then computed the JS divergence between their posteriors, to measure their similarity. 
We note that while eight parameters can be inferred for a input signal by \texttt{VItamin}, the three parameters which showed the greatest JS divergences were the component masses, $m_1$,$m_2$ and the merger time, $t_0$.

\begin{figure*} 
	\centering  
	\vspace{0.35cm} 
	\subfigtopskip=2pt 
	\subfigbottomskip=0.5pt 
	\subfigcapskip=-5pt
	\subfigure[]{
		\label{plt:jsd_ref_t0}
		\includegraphics[width=\columnwidth]{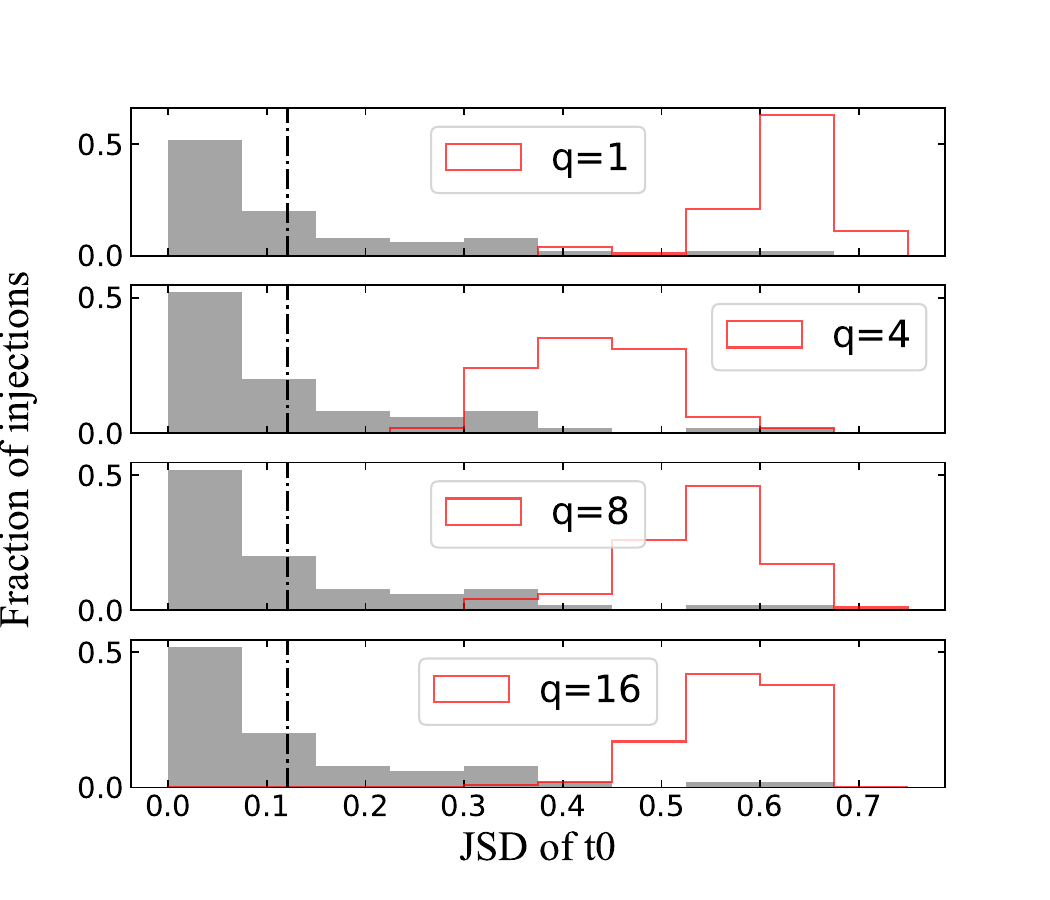}}
	\subfigure[]{
		\label{plt:jsd_ref_m1}
		\includegraphics[width=\columnwidth]{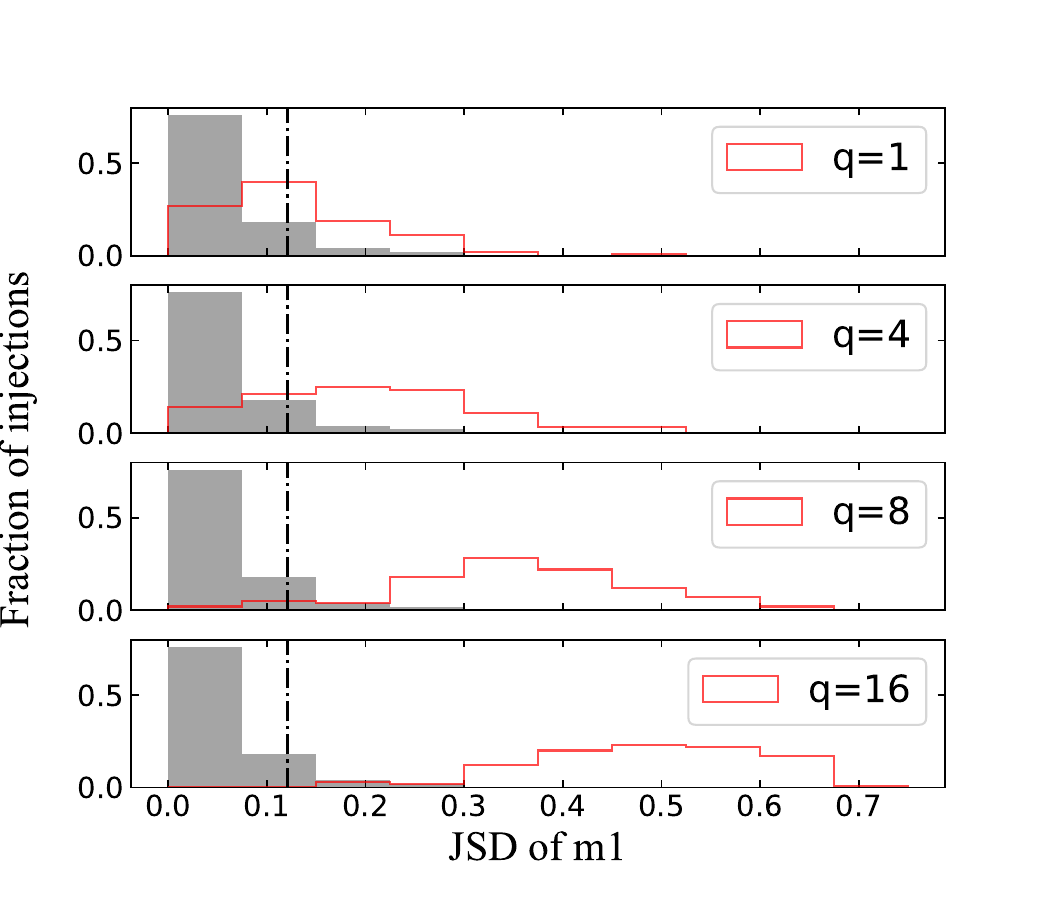}}
	\subfigure[]{
		\label{plt:jsd_ref_m2}
		\includegraphics[width=\columnwidth]{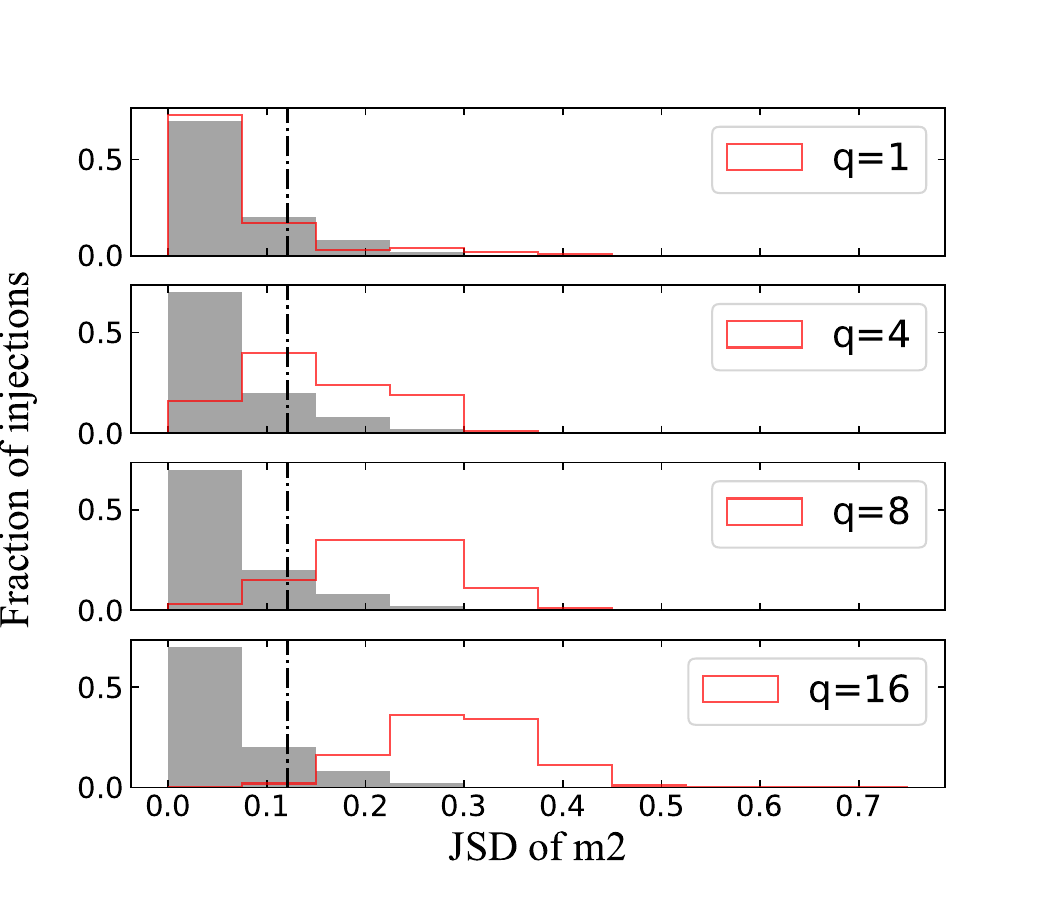}}
	\caption{
	Distributions of the JS divergence between parabolic BH capture and reference BBH $\jsref$ shown as the outline histograms and the JS divergence between reference BBHs with different noise $\jsnoise$, shown as the shaded histogram.
	The JS divergence analysis is performed for four waveforms with mass ratio $q$ = 1, 4, 8, 16 and three parameters $m_1$, $m_2$, and $t_0$ respectively.
	We evaluated our distinguishing method in terms of stability and effectiveness. 
	The former is illustrated by very low distributions of $\jsnoise$, which have 90\ \% upper limit, shown as a dashed line, of $m_1$, $m_2$, $t_0$ at 0.121, 0.134, 0.309. The latter is demonstrated by a high gap between distributions of $\jsnoise$ and $\jsref$, especially regarding JS divergence of $t_0$. The exact information about it is presented in Table~\ref{tab:90confidence}. This demonstrates that our approach works well.}
	\label{plt:jsd_ref}

\end{figure*}

Having computed the JS divergence for all 100 pairs of signals, we looked at the distribution of the divergences in Figure~\ref{plt:jsd_ref}, where the three subplots represent the JS divergences of the components masses $m_1$, $m_2$, and the merger time $t_0$.
The distribution of $\jsnoise$ is generally close to zero, suggesting that the effect of noise on \texttt{VItamin}'s posterior is rather limited as we hope.
We calculated $D_{90}$, the 90\% confidence interval of $p(\jsnoise)$, and used this as a threshold.
Then the percentage of $\jsref$ which is higher this threshold can indicate how far the distribution of $\jsref$ is away from the noise benchmark.
The related result is recorded in Table~\ref{tab:90confidence}.
The $D_{90}$ are 0.121, 0.134, 0.309 for $m_1$, $m_2$, $t_0$ respectively.
Though $D_{90}$ of $t_0$ is noticeably larger than those of the component masses of high, there is a large gap between distributions of $\jsnoise$ and $\jsref$ for this parameter, the percentage of which reaches 100\ \% for three waveforms and 97\ \% for the other one.
The greatest difference between the posteriors comes from the bias of the BBH approximatant \texttt{IMRPhenomPv2}.
For the same injection $t_0$ = 0.22\ s, a BBH signal is recovered with a peak value of 0.22\ s, but a parabolic BH capture is more likely to be recovered slightly later with, a recovered peak value of 0.25\ s or 0.26\ s (See this in Table~\ref{tab:bias}).
Therefore, this bias can be demonstrated through the the JS divergence analysis and used to test if a signal is a parabolic BH capture.
Besides, we also find that, for $m_1$, the average percentage of $\jsref$ above $D_{90}$ is 79.5\ \%, which has a more discriminative effect than that of $m_2$.
Parabolic captures with mass ratio of 8 and 16 can be distinguished fairly well from BBH signals, and the lowest percentage of them that higher than the threshold is also as high as 85\ \%.
This means we could have great confidence to distinguish the two types of signals when analysing with a BBH waveform. 

However, under a more realistic detection scenario, we have no access to the true parameters of the signal. Thus, in addition to calculating $\jsref$, we should also compare the posteriors of the parabolic BH capture and its recovered signal, and look for the evidence of the bias. 
Therefore, the recovered peak values were taken the average from 100 samples and used to inject the non-spinning BBH model \texttt{IMRPhenomPv2} with the same noise realisation.
The injections are recorded in Table~\ref{tab:bias}.
The new JS divergence we introduce is: 
\begin{enumerate}
\item $\jsrec$: the JS divergence between the posterior of a parabolic BH capture and its recovered BBH signal injected using the recovered peak values, with the same noise realisation.
\end{enumerate}

$\jsrec$ describes the effect of recovering parabolic BH capture.
If the input signal is actually a BBH merger, its recovered signal should have a very similar posterior probability density distribution.
In this case, the distribution of $\jsrec$ is close to zero but slightly higher due to the noise, of which the effect can be represented by $\jsnoise$.
But for other signals, if the difference between the posteriors of the recovered signal and itself is great, then $\jsrec$ could be used as a criterion. 

\begin{figure*} 
	\centering  
	\vspace{0.35cm} 
	\subfigtopskip=2pt 
	\subfigbottomskip=0.5pt 
	\subfigcapskip=-5pt 
	\subfigure[]{
		\label{plt:jsd_inj_t0}
		\includegraphics[width=\columnwidth]{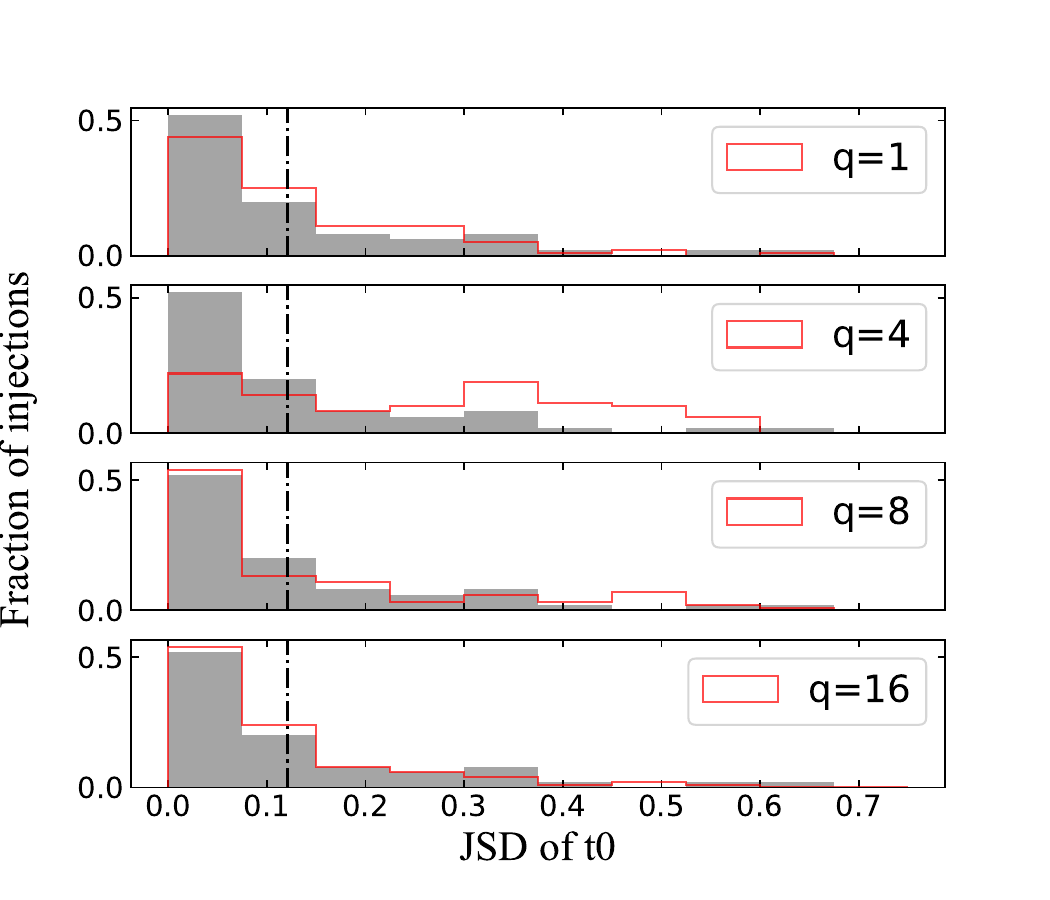}}
	\subfigure[]{
		\label{plt:jsd_inj_m1}
		\includegraphics[width=\columnwidth]{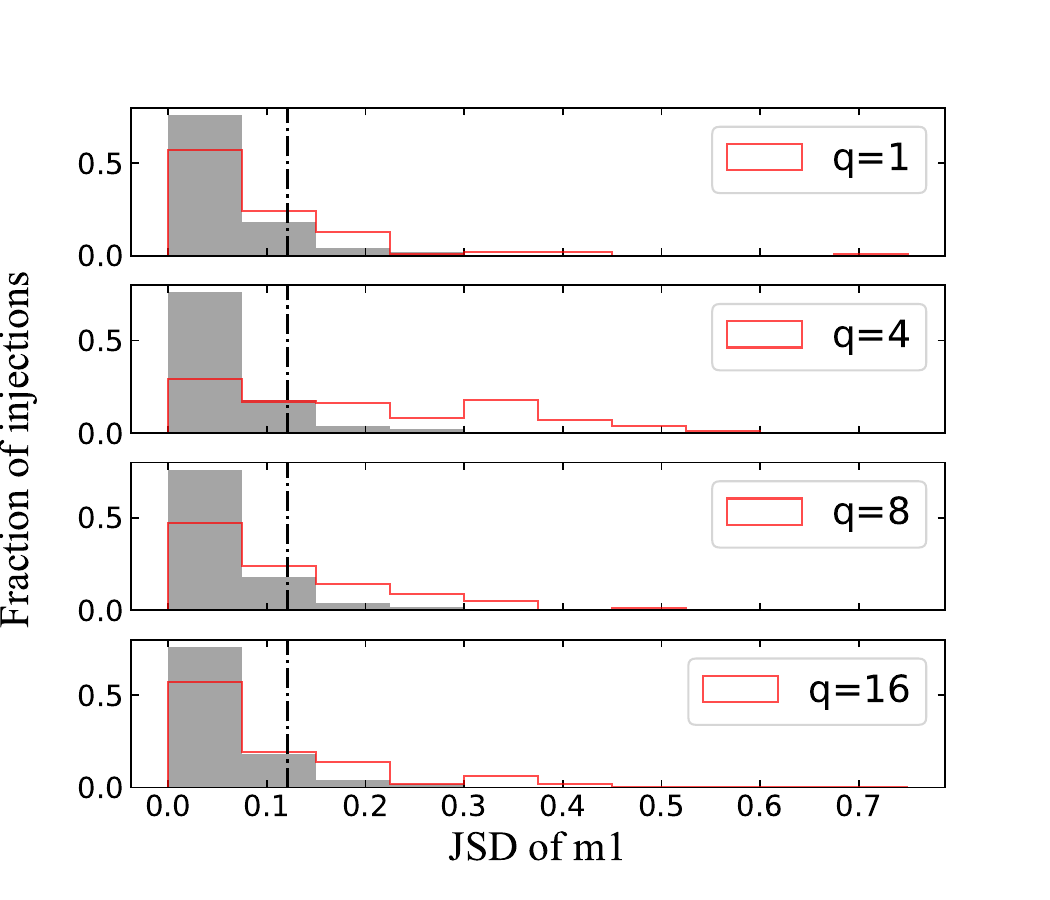}}
	\subfigure[]{
		\label{plt:jsd_inj_m2}
		\includegraphics[width=\columnwidth]{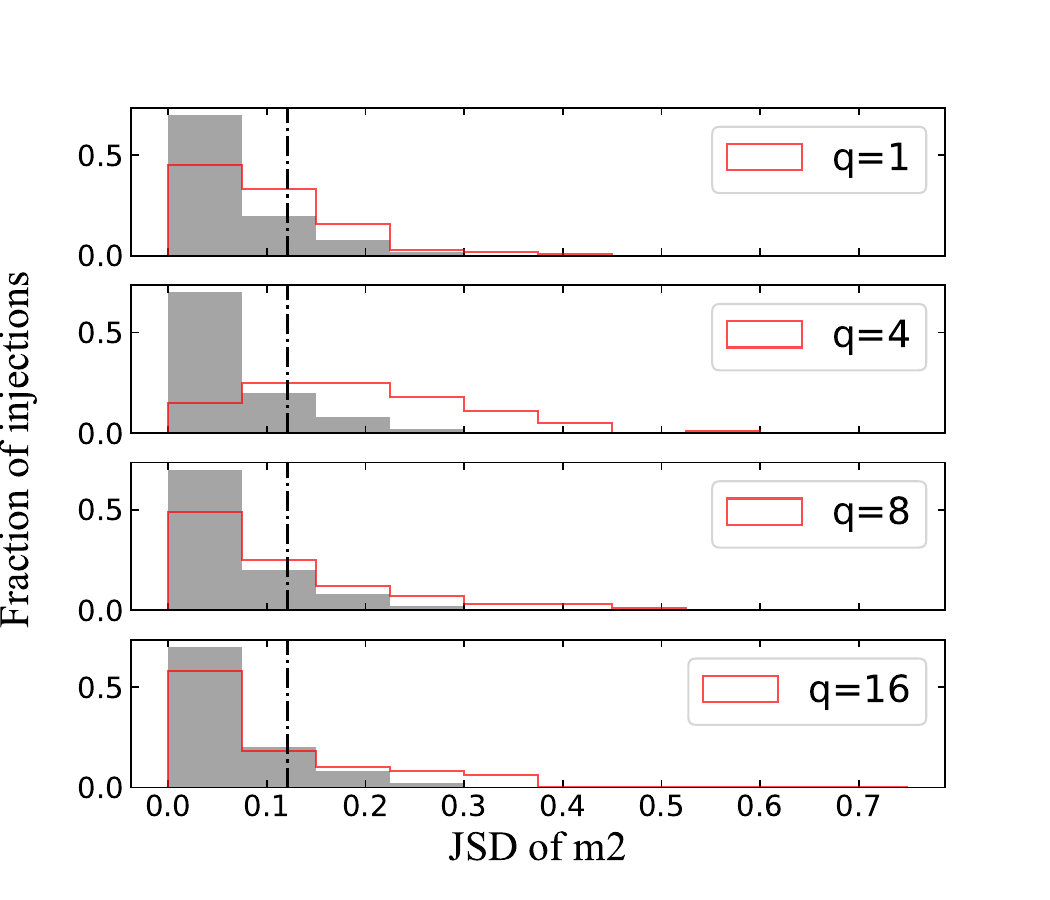}}	
	\caption{Distributions of the JS divergence between parabolic BH capture and its recovered BBH $\jsrec$ shown as the outline histograms, and the JS divergence between reference BBHs with different noise $\jsnoise$ shown as a shaded histogram. The JS divergence analysis is performed for four waveforms with mass ratio $q$ = 1, 4, 8, 16 and three parameters $m_1$, $m_2$, and $t_0$ respectively. 
	Here we considered the application of our distinguishing method in more realistic scenarios and evaluated it in terms of stability and effectiveness. The stability is the same as before with very low distributions of $\jsnoise$, with the 90\ \% upper limit represented by a dashed line. However, the distributions of $\jsnoise$ and $\jsrec$ almost overlap which suggests that the two types of signal can't be well distinguished in this situation.(More information about this is presented in Table~\ref{tab:90confidence}.)}
	\label{plt:jsd_inj}

\end{figure*}

We plotted the distributions of $\jsrec$ and compared it with the noise benchmark in Figure~\ref{plt:jsd_inj}. 
Three subplots represents JS divergence of $m_1$, $m_2$, and $t_0$. 
However, it almost overlaps with the distributions of $\jsnoise$ for the three parameters, which suggests a high similarity between posteriors of parabolic BH capture and the recovered high-mass BBH. 
We also determined the percentage of $\jsrec$ which were higher than $D_{90}$ and recorded them in Table~\ref{tab:90confidence}. 
$\jsrec$ has a much lower percentage than $\jsref$ above $D_{90}$. 
Except for the waveform with mass ratio of 4, the highest percentage of their $\jsrec$ greater than the threshold is only $32\ \%$.
The waveform with mass ratio of 4 is unusual compared to others, with the average percentage reaching $57.3\ \%$. Since we do not know \textit{a priori} what the mass ratio of the waveform is in the real analysis, we must consider all the waveforms equally, so this value would not be enough to provide support that the evidence of the bias has been found. 
In addition, $\jsref$ has a good performance on $t_0$, while $\jsrec$ is difficult to tell apart from $\jsnoise$ as it has an average percentage of $20.5\ \%$ above $D_{90}$. 

\begin{table}
\centering
\begin{tabular}{ccccclll}
\cline{1-5}
\multicolumn{1}{l|}{} & \multicolumn{1}{c|}{mass ratio} & $t_0$ & $m_1$ & $m_2$ & & &  \\ \cline{1-5}
\multicolumn{1}{c|}{\multirow{4}{*}{$\jsref$}} & \multicolumn{1}{c|}{1} & 100\% & 49\% & 11\% &  &  &  \\
\multicolumn{1}{c|}{} & \multicolumn{1}{c|}{4} & 97\% & 74\% & 52\% & & & \\
\multicolumn{1}{c|}{} & \multicolumn{1}{c|}{8} & 100\% & 95\% & 86\% & & & \\
\multicolumn{1}{c|}{} & \multicolumn{1}{c|}{16} & 100\% & 100\% & 100\% & & & \\
\multicolumn{1}{c|}{\multirow{4}{*}{$\jsrec$}} & \multicolumn{1}{c|}{1} & 9\% & 28\% & 24\% & & & \\
\multicolumn{1}{c|}{} & \multicolumn{1}{c|}{4} & 46\% & 62\% & 64\% & & & \\
\multicolumn{1}{c|}{} & \multicolumn{1}{c|}{8} & 19\% & 32\% & 27\% & & & \\
\multicolumn{1}{c|}{} & \multicolumn{1}{c|}{16} & 8\% & 32\% & 27\% & & & \\ \cline{1-5}

\end{tabular}
\caption{The percentage of $\jsref$ and $\jsrec$ higher than the noise threshold for parabolic BH capture waveform with mass ratio of $1, 4, 8$, and $16$. The threshold is represented by $\jsnoise$ at $90\ \%$ confidence level, which is $0.309, 0.121, 0.134$ for $t_0$, $m_1$, and $m_2$ respectively.}\label{tab:90confidence}
\end{table}

Apart from being used for JS divergence analysis, Table~\ref{tab:bias} also gives us inspiration about the patterns on injected and recovered parameters. First, for parabolic BH capture waveforms with mass ratios of $1, 4, 8$, and $16$, the total mass is recovered as $144, 163, 181$, and $194$ $\msun$ respectively. These amount to a tendency for the rising of the recovered total mass with the mass ratio $q$ increasing, and the former one is much higher than the injection of $150$ $\msun$ when the mass ratio is greater than $1$. 
We also find that, the recovered mass ratios are $1.12, 1.17,1.18,$ and $1.16$ respectively, which are all close to one. 
For comparison, GW190521 has a mass ratio of $1.29$, and it is basically consistent with the analysis result we got. 
The sensitive distance decreases with the increase in the mass ratio $q$.
For equal-mass BH binaries, eccentric sources are thought to be much closer than BBH sources with a circular orbit in inspiral. 
Another discrepancy that could be highlighted is that the recovered merger times $t_0$ are all about 0.04 $s$ behind the injection truth. We suspect that it is caused by a mathematical fit of the BBH model to the capture signal, but we will investigate for a deeper pattern in the future.  

\section{Summary and discussion}
\label{sec:conclusion} 

In this work, we proposed the possibility that current approaches to GW analysis could  misclassify parabolic BH capture signal as a BBH signal.
We then demonstrated a scenario under which this could occur, and devised for a statistical method to distinguish them. 
We injected parabolic BH capture waveforms to produce mock data, using a tool developed for characterising burst searches, \texttt{Minke}, which was exploited to make injection with the customized distribution. 
The main difficulty is that, it is impossible to predict how a signal be inferred under a biased multi-parameter model, and the computational cost of traditional Bayesian inference is expensive.
To overcome this we adopted \texttt{VItamin}, a neural network based on the BBH model, and retrained it to fit high-mass BBH signals, which reduced the cost of each parameter estimation to a very low level.
This greatly helped us to continuously adjust the injection parameters of the parabolic BH capture and finally obtain the appropriate posterior probability. 
After that, we also performed confirmatory parameter estimation using \texttt{dynesty} sampler, of which the result also had a strong statistical support. 

Here we summarize our main conclusions in more detail. 

We have established that there are scenarios in which a parabolic BH capture could be recovered as a spinning (non-spinning) BBH signal with high statistical support, a log Bayes factor of $\ln{K}=134.6\ (111.7)$, compared to a noise hypothesis.
This type of signal is likely to be mistaken as a high-mass BBH by LIGO and Virgo.
Therefore it would be valuable to be vigilant to this possibility when a high-mass BBH system is identified in an analysis, otherwise future GW events may be misclassified.
This should be considered in cases where the waveform seems to lack a clear inspiral phase. 

In this study, we have built a rapid approach to describe the difference between the posteriors of BBH and parabolic BH capture signals and distinguish them. 
This approach is based on neural network, \texttt{VItamin}, and compares the distribution of JS divergences of three parameters $m_1$, $m_2$, and $t_0$ from two types of GW signals, with that of noise benchmark $D_{90}$. 
Its validity has been proved by the JS divergence between the parabolic BH capture and the reference BBH, $\jsref$, which has 79.5\ \%, 62.3\ \%, 99.3\ \% of samples over $D_{90}$ for $m_1$, $m_2$, and $t_0$. 
However, in a more realistic detection scenario, our analysis does not yield evidence that two types of GW events are distinguishable with the current BBH Bayesian inference. 
This is a result of the lower value of the JS divergence between the parabolic BH capture and its recovered BBH $\jsrec$, containing only 38.5\ \%, 35.5\ \%, 20.5\ \% of samples located above $D_{90}$ for $m_1$, $m_2$, and $t_0$. 
The result of our analysis would not therefore allow us to make an identification of a GW190521-like signal.
As a result the parabolic BH capture could not be distinguished from a BBH by the current quasi-circular BBH analysis, which highlights the importance of a good BH capture approximant in the future. 

We have identified the patterns on injected and recovered parameters. For four waveforms, there is a tendency for the recovered total mass to rise as the mass ratio increases; only the one from equal-mass system has a recovered total mass close to the injection of 150 $\rm{M}_{\odot}$, and the total masses of the others are recovered with much higher values. The recovered mass ratios are all close to one, which we also see on GW190521 with a mass ratio of 1.29. In contrast to the pattern observed with the total mass, the sensitive distance decreases as the mass ratio increases. We also note that that the recovered merger times are all offset by around 0.04 s compared to the injected value.

The research in this paper constitutes a comparatively novel use of deep learning in GW data analysis.
A typical Bayesian approach to analyses used in this study takes 8 to 14 hours while the neural network requires around $50$ seconds. 
For each waveform, there were about four iterations on average before determining the appropriate $d_{\rm L}$ injection, and each turn gave 100 Bayesian posteriors corresponding to the combinations of sky location. 
A total of $1,600$ inferences were performed in this stage. 
Once the posteriors which mimic BBH were obtained, we selected one signal from each waveform and analysed it with 100 noise realisations, as well as the reference BBH signal, for construction of JS divergence distribution, of which the stage contained 500 inferences. 
BBH signals injected from the recovered peaks of the BH capture signals were then inferred with the same noise realization sets. 
This last step required 400 inferences and constructed the distribution of $\jsrec$ to finally describe the difference between BBH and BH capture signals. 
Overall, the use of a neural network saved around $2.7\times 10^4$ hours when performing $2,500$ parameter estimation analyses.  

Because of computational cost limitations in training, the \texttt{VItamin} network has not been trained to take into account the spins of the BBH model.
One promising signature of the BH binary formation environment is the angular distribution of BH spins~\citep{distinguish_spins}. 
Binaries formed through dynamical interactions are expected to have isotropic spin orientations~\citep{iso_spins_1,iso_spins_2,iso_spins_3,iso_spins_4,iso_spins_5} whereas systems formed from pairs of stars born together are more likely to have spins preferentially aligned with the binary orbital angular momentum~\citep{aligned_spins_2,aligned_spins_3,aligned_spins_4,aligned_spins_5}.
When modeling the BH capture data, the six additional parameters of spins, as intrinsic properties of a binary, are expected to play an important role in distinguishing binaries formation channel, allowing a further precise search that has been done in the real data analysis.
We will return to this subject in future work.

The component masses prior range of \texttt{VItamin} can be expanded and the sampling rate can be raised to cover more BH capture samples. 
These events are principally from low-frequency sources, and as such are ideal candidates for both Einstein Telescope~\citep{ET_1,ET_2}, which aims to achieve much greater low-frequency sensitivity than current detectors, but also for Deci-Hz detectors, such as DECIGO~\citep{DECIGO_1,DECIGO_2}.
The misclassification is expected to be eliminated with their ability to observe at much lower frequencies, removing the ambiguity between unobserved low frequency inspiral cycles and a total lack of inspiral.
The detection rate of BH captures is dependent on the initial mass function of stars in galactic nuclei and the mass of the most massive BHs, therefore future observations can constrain both the average star formation properties and upper mass of BHs in galactic nuclei~\citep{bhencounter}.

\section*{Acknowledgements}

We would like to thank Charlie Hoy, Juan Calderon Bustillo, and Rossella Gamba for their comments on the manuscript, and suggestions, in addition to many discussions within the parameter estimation and burst groups of the LIGO, Virgo, and KAGRA collaborations.
DW and ISH were supported by STFC grants ST/V001736/1 and ST/V005634/1.
YB was supported by IBS under the project Code No. IBS-R018-D1 and by the National Research Foundation of Korea (NRF) grant funded by the Korea government(MSIT) (No. NRF-2021R1F1A1051269). 
GK and YB are supported by  the KISTI National Supercomputing Center with supercomputing resources and technical supports (KSC-2020-CRE-0352).
ZZ was supported by the National Natural Science Foundation of China under Grants Nos. 11633001, 11920101003 and 12021003, the Strategic Priority Research Program of the Chinese Academy of Sciences, Grant No. XDB23000000 and the Interdiscipline Research Funds of Beijing Normal University.
We are grateful for the support of our colleagues in the LIGO-Virgo Compact Binary Coalescence Parameter Estimation working group.
This work has been assigned LIGO document control number LIGO-P2200064.

\section*{Data Availability}
The data and code underlying this article are available in Zenodo at \url{https://doi.org/10.5281/zenodo.6384509}.
 



\bibliographystyle{mnras}
\bibliography{references} 




\appendix

\section{Full vitamin and bilby posteriors}

\begin{figure*} 
	\centering  
    \includegraphics[width=\linewidth]{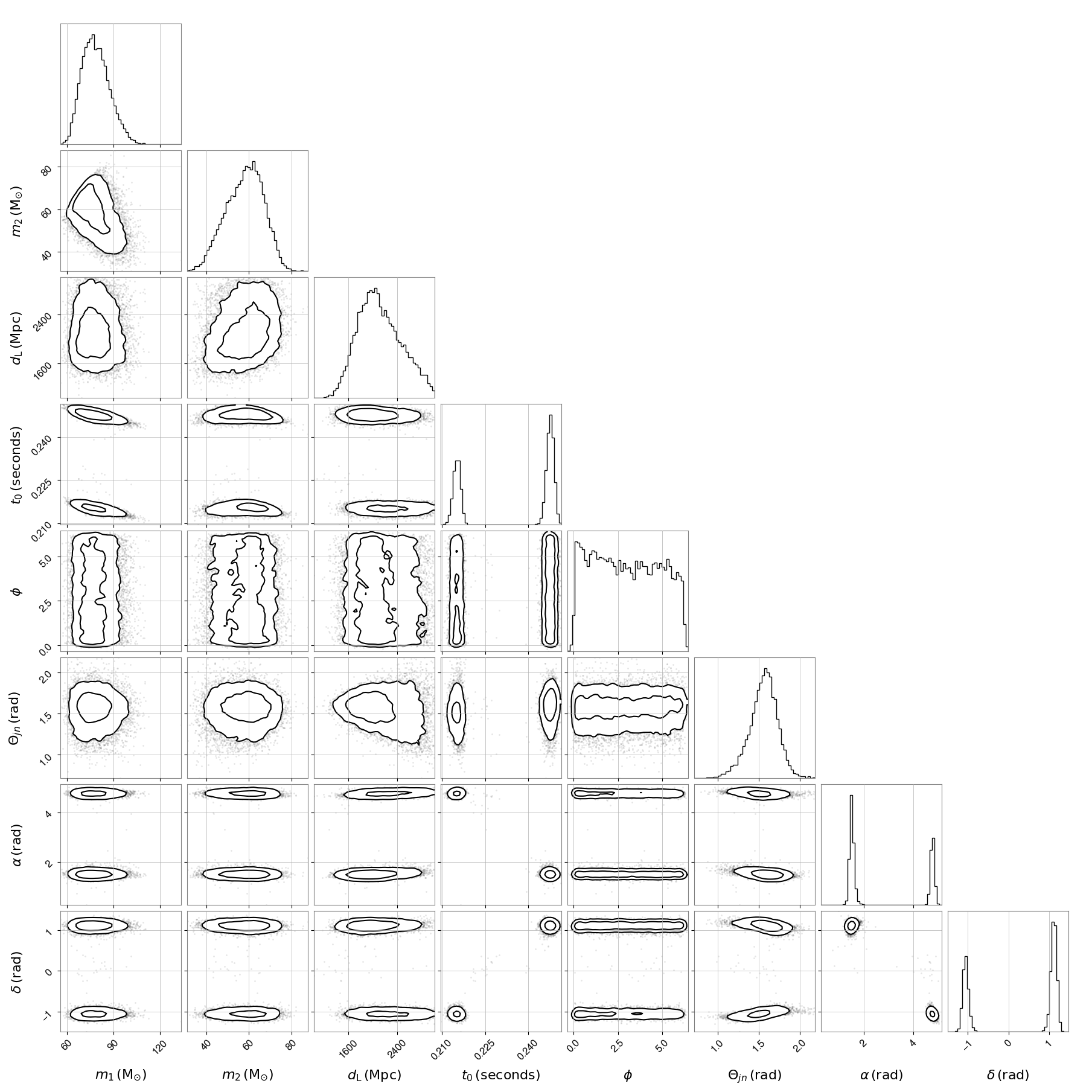}
	\caption{The posterior probability density distribution for a BH capture, recovered using a non-spinning BBH model and \texttt{VItamin}.}
	\label{vtm_be}
\end{figure*}

\begin{figure*} 
	\centering  
	\includegraphics[width=\linewidth]{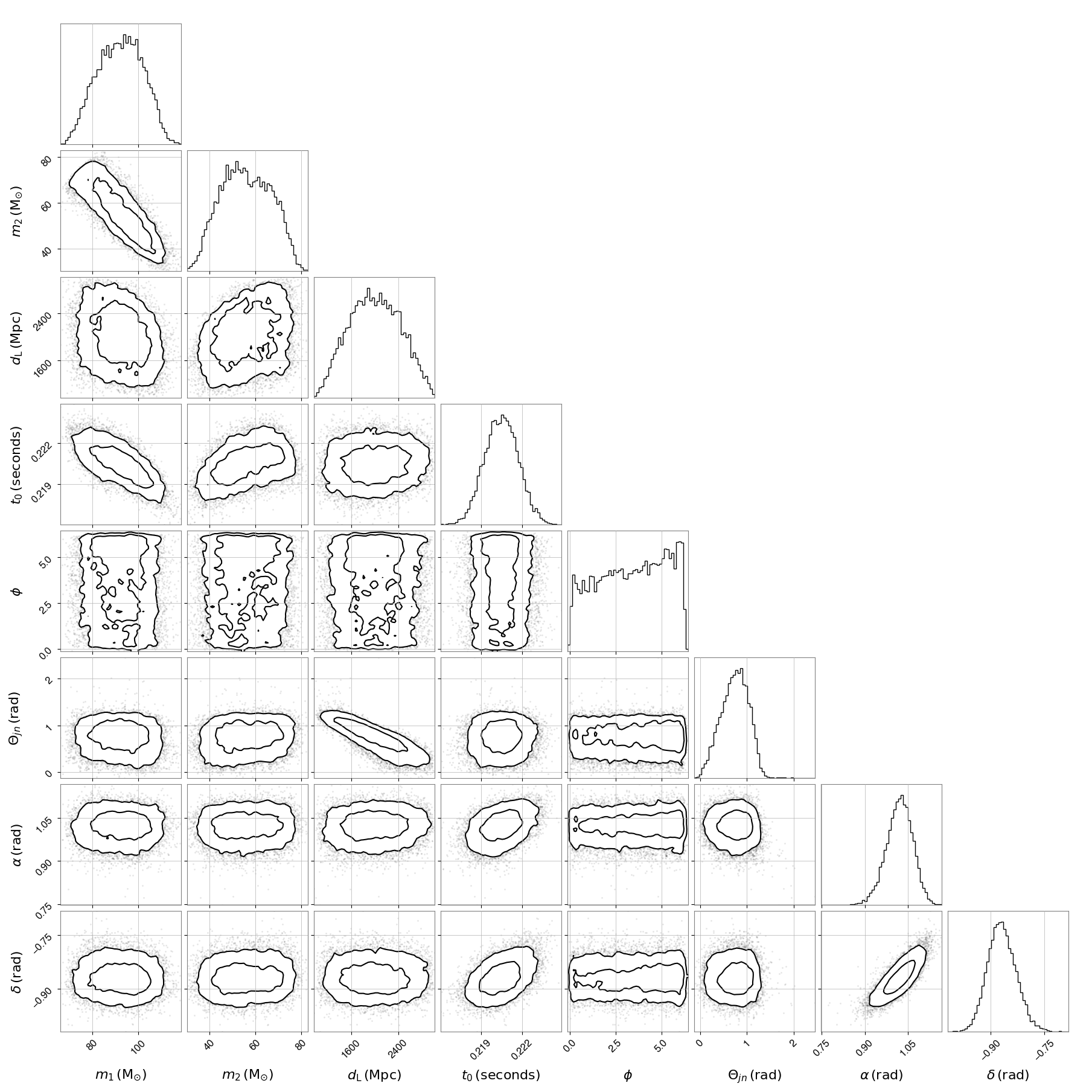}
    \caption{The posterior probability density distribution for a BBH, recovered using a non-spinning BBH model and \texttt{VItamin}.}
	\label{vtm_bbh}
\end{figure*}

\begin{figure*} 
	\centering  
        \includegraphics[width=\linewidth]{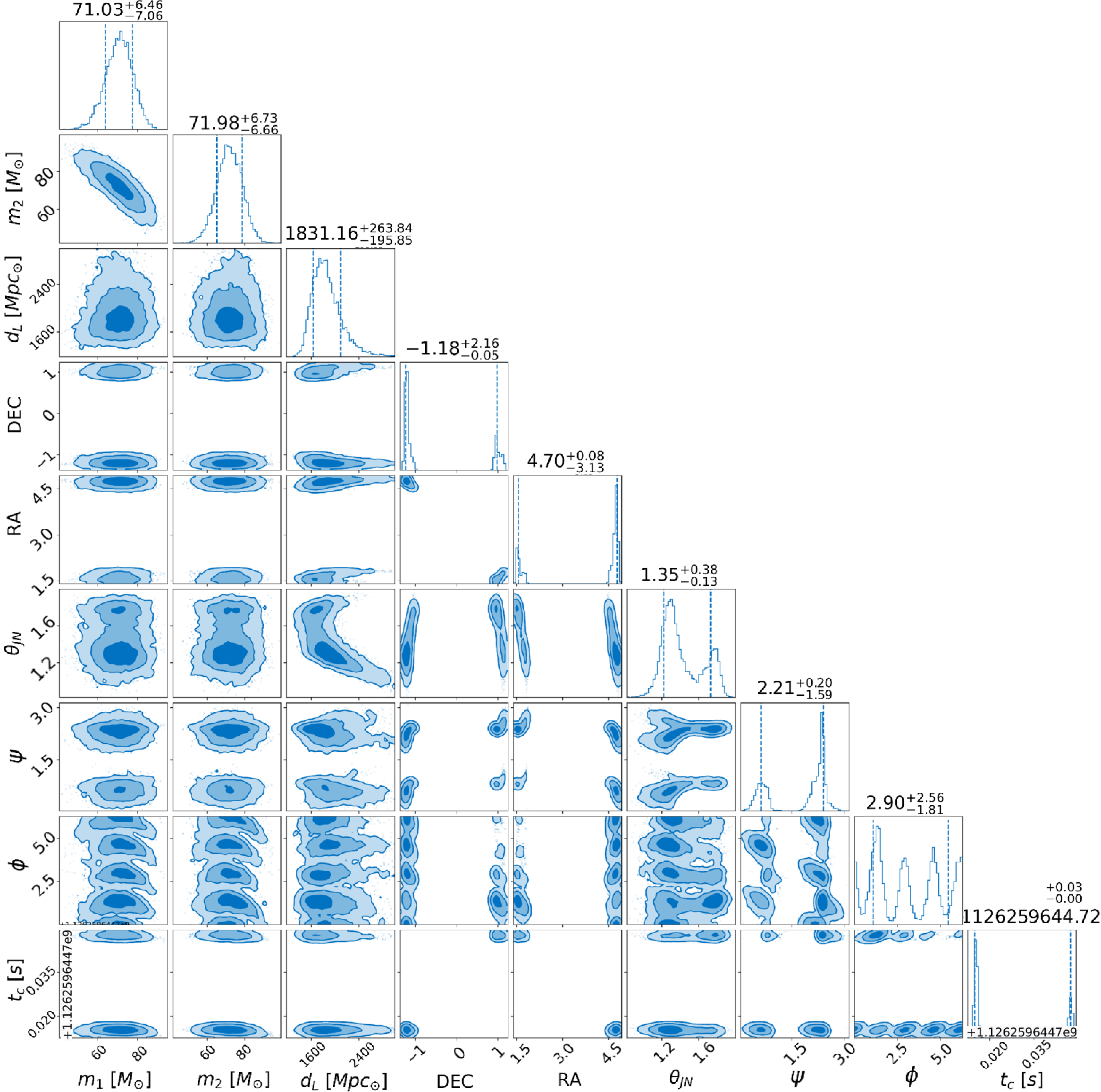}	
	\caption{The posterior probability density distribution for a BH capture, recovered using a non-spinning BBH model and the \texttt{dynesty} sampler.}
	\label{blb_be}
\end{figure*}

\begin{figure*} 
	\centering  
    \includegraphics[width=\linewidth]{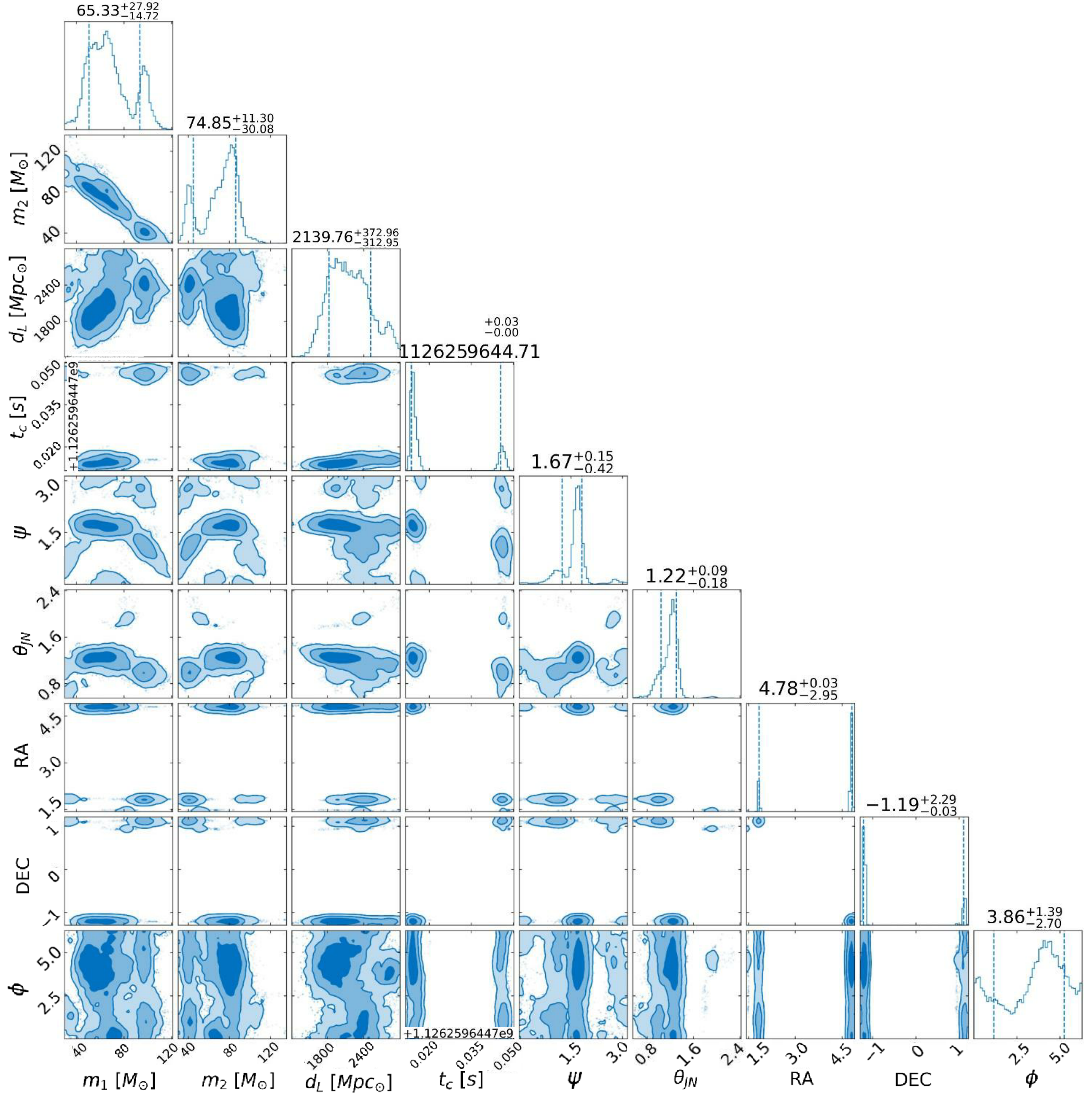}	
	\caption{The posterior probability density distribution for a parabolic BH capture, recovered using a spinning BBH model and the \texttt{dynesty} sampler. Here we show the posterior for the main parameters and omit 6 spins.}
	\label{blb_be_spin}
\end{figure*}

In this appendix we present corner plots for the posterior distributions produced by analysing a parabolic BH capture with both \texttt{VItamin}~(Figure~\ref{vtm_be}) and \texttt{Bilby} with a non-spinning BBH model (Figure~\ref{blb_be}), a binary black hole merger injection with \texttt{VItamin}~(Figure~\ref{vtm_bbh}) and \texttt{Bilby} with a spinning BBH model~(Figure~\ref{blb_be_spin}).

\bsp	
\label{lastpage}
\end{document}